# Predicted Abundances of Carbon Compounds in Volcanic Gases on Io

Laura Schaefer and Bruce Fegley, Jr.

*Planetary Chemistry Laboratory, McDonnell Center for the Space Sciences, Department of Earth and Planetary Sciences, Washington University in St. Louis*

*e-mail:* laura_s@levee.wustl.edu *and* bfegley@levee.wustl.edu

## ABSTRACT

We use chemical equilibrium calculations to model the speciation of carbon in volcanic gases on Io. The calculations cover wide temperature (500–2000 K), pressure ($10^{-8}$ to $10^{+2}$ bars), and composition ranges (bulk O/S atomic ratios ~0 to 3), which overlap the nominal conditions at Pele (1760 K, 0.01 bar, O/S ~ 1.5). Bulk C/S atomic ratios ranging from $10^{-6}$ to $10^{-1}$ in volcanic gases are used with a nominal value of $10^{-3}$ based upon upper limits from Voyager for carbon in the Loki plume on Io. Carbon monoxide and $CO_2$ are the two major carbon gases under all conditions studied. Carbonyl sulfide and $CS_2$ are orders of magnitude less abundant. Consideration of different loss processes (photolysis, condensation, kinetic reactions in the plume) indicates that photolysis is probably the major loss process for all gases. Both CO and $CO_2$ should be observable in volcanic plumes and in Io's atmosphere at abundances of several hundred parts per million by volume for a bulk C/S ratio of $10^{-3}$.

## 1. INTRODUCTION

Galileo, Hubble Space Telescope (HST), and Earth-based observations show high temperature silicate volcanism on Io, the most volcanically active object in the solar system (Kargel et al. 2003; Lopes et al. 2001; Spencer & Schneider 1996). Volcanic gases on Io are composed of sulfur and oxygen compounds ($SO_2$, S, $S_2$, SO) with smaller amounts of sodium, chlorine, and potassium compounds such as Na, K, and NaCl (Fegley & Zolotov 2000a; Lellouch et al. 2003; McGrath et al. 2000; Spencer et al. 2000; Zolotov & Fegley 1998, 1999, 2001). A comparison of observations with thermodynamic and kinetic calculations shows that volcanic gases on Io, as on Earth, are in chemical equilibrium when they erupt (Zolotov & Fegley 1998, 1999, 2001, Fegley & Zolotov 2000a, b). Terrestrial volcanic gases are mainly $H_2O$, $SO_2$, and $CO_2$ with smaller amounts of $H_2S$, CO, $H_2$, OCS, HCl, and HF (e.g., see Table 6.13 of Lodders & Fegley 1998). Io is depleted in hydrogen, and H-bearing gases such as $H_2O$, $H_2S$, and $H_2$ are not expected in Ionian volcanic gases (Fegley & Zolotov 2000a). But what about carbon gases such as $CO_2$, CO, and OCS?

There are good reasons to suspect that carbon is present on Io and here we use chemical equilibrium and kinetic calculations to predict carbon chemistry in Ionian volcanic gases. First, Voyager upper limits for C-bearing gases in the Loki plume (Pearl et al. 1979) are fairly high (0.02–3% relative to $SO_2$). Second, the Loki plume observed by Voyager was apparently generated by interaction of a lava flow with sulfur-rich surface material (Geissler 2003). The carbon to sulfur ratio in this plume may be lower than in volcanic gases evolved at Pele or other hot spots. Third, carbon to sulfur atomic ratios are a few percent or higher (Fig. 1) in lunar rocks, basaltic meteorites, and other samples depleted in volatiles relative to solar abundances. It is plausible that carbon remains in the silicate portion of Io and can be volcanically outgassed. Our chemical equilibrium calculations are described in § 2. We discuss constraints on Io's carbon abundance in the next section (§ 3). This is followed by the results of our chemical equilibrium calculations for the carbon chemistry of Ionian volcanic gases (§ 4). We then use chemical kinetic calculations to model disequilibrium chemistry in the volcanic plume and find that photolysis is the major loss process for carbon-bearing gases (§ 5). Conclusions are given in § 6. Preliminary results of this work are reported by Fegley & Zolotov (2000b).

## 2. EQUILIBRIUM CALCULATIONS

We computed the chemical equilibrium abundances of gaseous (i.e., atoms, radicals, ions, electrons, and molecules) species in Ionian volcanic gases containing the elements S, O, Na, K, Cl, and C. The abundance and stability of liquids and solids that may condense from the gas were also computed. The calculations were done using a Gibbs energy minimization code of the type described by Van Zeggern & Storey (1970) and the CONDOR mass-balance, mass-action code described by Fegley & Lodders (1994). The two codes give identical results.

Computational methods and thermodynamic data are described elsewhere (Fegley & Lodders 1994, Fegley & Zolotov 2000a).

As in our prior work we did calculations over a range of temperatures (500–2000 K), pressures ($10^{-8}$ to $10^{+2}$ bars), and bulk O/S atomic ratios (~0 to 3). We use the Pele volcanic vent as a nominal case because it is the best characterized volcanic vent on Io. Galileo and Cassini observations indicate that temperatures of the Pele volcanic vent range from about 1400 K to 1760 K (Radebaugh et al. 2004; Lopes et al. 2001). We use a nominal temperature of T = 1760 ± 210 K from Galileo NIMS observations (Lopes et al. 2001). Zolotov & Fegley (2001) determined the vent pressure as 0.01 bar using chemical equilibrium calculations. As we show later, our key conclusions are relatively insensitive to temperature and pressure over a fairly wide range of plausible values.

We used the same relative atomic abundances for S (1.0), O (1.521), Na (0.05), K (0.005), and Cl (0.045) as in our previous modeling (Schaefer & Fegley 2004, Fegley & Zolotov 2000a, b, Zolotov & Fegley 2001, Moses et al. 2002). These elemental abundances are based upon spectroscopic observations of the Pele plume, Io's extended atmosphere, and the Io plasma torus (IPT). Carbon has not been detected on Io and the carbon elemental abundance is a variable in our calculations, which we discuss in the next section.

## 3. CONSTRAINTS ON IO'S CARBON ABUNDANCE

We use three data sets to constrain the abundance of carbon in Io: (1) geochemical analyses of carbon and sulfur in stony meteorites, lunar, and terrestrial samples, (2) spectroscopic upper limits on carbon-bearing gases in the Loki and Pele volcanic plumes on Io, and (3) spectroscopic upper limits on ionized carbon in the IPT. All data are plotted in Fig. 1.

The C/S ratios measured in meteorites are from the METBASE meteorite database (Koblitz 2003) and are for observed meteorite falls. The lunar data are for soils, breccias, and basalts (Fegley & Swindle 1993). The terrestrial data are for the upper and lower continental crust, present day mantle, and different rock types including basalt, kimberlite, granite, shale, and ultramafic rocks (Lodders & Fegley 1998).

Chondrites are undifferentiated stony meteorites that contain metal, silicates, and sulfides. They were not melted on their parent bodies. The C/S ratios of chondrites range from 0.011 – 4.27. Carbonaceous chondrites have the highest C/S ratios and ordinary chondrites the lowest ones. Achondrites are differentiated stony meteorites that have lost essentially all metal and sulfides during melting on their parent bodies. The different types of achondrites for which analytical data are available have C/S ratios ranging from 0.10 – 25.4. The highest C/S ratios are found in ureilites, which are carbon-rich and contain diamonds and organic compounds, whereas the lowest C/S ratios are found in the enstatite achondrites (aubrites), mainly composed of enstatite ($MgSiO_3$). The lunar samples have C/S ratios from 0.05 – 0.6. The highest C/S ratios are for soils while the lowest are for basalts. Finally, the terrestrial samples have C/S ratios ranging from 0.07 for ultramafic rocks to 21.6 for diamond-bearing kimberlites.

Infrared spectroscopy by Voyager gave upper limits for the abundances of several carbon-bearing gases in the Loki plume (Pearl et al. 1979). They reported 100 nanobars (nbar) $SO_2$ in the Loki plume and give two different upper limits for other gases corresponding to gas temperatures of 130 and 250 K. Following Zolotov and Fegley (1999) we use the arithmetic mean of the upper limits for each gas. The results expressed as volume (or mole) percentages relative to $SO_2$ are $CO_2$ (< 0.05%), OCS (< 0.1%), $CS_2$ (< 0.02%), and $CH_4$ (< 3%).

Lellouch et al. (1992) reanalyzed the Voyager data and derived lower $SO_2$ abundances (5–40 nbar) as a function of the assumed gas temperature. If we combined their lower $SO_2$ abundances with the upper limits (cm-am) reported by Pearl et al. (1979) the $CO_2/SO_2$, $OCS/SO_2$, $CS_2/SO_2$, and $CH_4/SO_2$ ratios would become larger than the values we give above. However, since Lellouch et al. (1992) did not re-determine the upper limits for the carbon-bearing gases, we use the values based on the $SO_2$ abundance of 100 nbar (0.2 cm-am) from Pearl et al. (1979), rather than combine abundances determined by separate analyses. This is a reasonable action because we are looking at relative rather than absolute abundances.

Ultraviolet spectroscopy from the HST (McGrath et al. 2000) gives upper limits for $CS_2$ in volcanic plumes at Pele (< 0.64% of $SO_2$) and Ra (< 1.33 % of $SO_2$). As our calculations will show, $CS_2$ is a very minor C-bearing gas, so these upper limits do not significantly constrain the bulk atomic C/S ratio at Pele or Ra. Earth-based millimeter wave spectroscopy gave upper limits $< (0.6\text{–}3) \times 10^{-2}$ cm-am for CO in Io's global atmosphere (Lellouch et al. 1992). This corresponds to about 30–150% of $SO_2$ using the typical $SO_2$ abundance of 0.02 cm-am given by Lellouch et al. (1992). Finally, short wavelength UV observations of the IPT by Feldman et al. (2004) give upper limits of $< 3.7\times10^{-4}$ for the $C^{2+}/S^{2+}$ atomic ratio and $< 2.5\times10^{-3}$ for the $C^{+}/S^{2+}$ ratio. With the exception of the millimeter wave upper limit for CO in Io's global atmosphere, which is clearly too large, the other spectroscopic upper limits for carbon-bearing gases on Io and for ionized carbon in the Io plasma torus are shown together (as the bar labeled upper limits on Io) in Fig. 1.

The referee noted that if all upper limits are good quality, only the most stringent one is relevant. This would be the upper limit on C/S ions in the IPT. However, there are a number of different processes (e.g. ionization, atmospheric sputtering, and molecular diffusion) which affect the composition of the IPT (Spencer & Schneider 1996), so these values may not be strictly applicable to the bulk atomic C/S ratios in volcanic gases. Additionally, as noted above, we have upper limits for C-bearing gases at several different hot spots. There is no reason to assume that the lava at each of these hot spots has the same bulk C/S ratio (see for instance the range of values for terrestrial samples). We therefore compare the range of upper limits to the values for terrestrial, lunar and meteoritic samples below.

Taken at face value, the upper limits for Io are significantly smaller than the C/S ratios for meteoritic, lunar, and terrestrial samples. This is somewhat surprising because the Moon and the basaltic achondrites (eucrites) are generally depleted in volatiles (e.g., $H_2O$, C, N, S) relative to CI chondritic abundances, and these objects should be good analogs for Io. In particular, Io has about the same size, mass, and density as the Moon, and both objects are bone-dry.

One could argue that Io lost its carbon by volcanic outgassing, while the Moon, which is geologically quiescent, retained its carbon. However, the giant impact model for the formation of the Moon predicts that lunar-forming material was heated to very high temperatures of several thousand kelvins (e.g., Canup 2004). Yet the Moon still contains carbon with about 20–30 μg/g carbon in basalts and larger concentrations of carbon in other rock types and soils (Fegley & Swindle 1993). Also, the loss of carbon and sulfur via outgassing on Io may be similar because the molecular masses of carbon and sulfur gases that are observed to be outgassed (SO, $S_2$, $SO_2$) on Io or Earth (CO, $CO_2$, OCS) are similar within a factor of 2.3. Thus, it seems difficult to ascribe the apparent lack of carbon on Io to its high temperature volcanism.

Alternatively, as the referee suggested, Io's lithosphere and surface may be enriched in sulfur, rather than depleted in carbon. This seems plausible given the evidence for sulfur compounds in Io's atmosphere, volcanic gases, and surface. This is especially true for volcanic plumes which are formed by interaction of hot lavas with sulfur-rich surface substrate, such as the plume observed by Voyager at Loki (Geissler 2003) and the persistent plume observed at Prometheus (Kieffer et al. 2000, Milazzo et al. 2001). Vaporization of the surface frost will therefore lower the bulk atomic C/S ratios in some volcanic gases. However, plumes at other hot spots such as Pele are directly exsolved from the hot lava (Spencer et al. 1997), and so should be less contaminated by the sulfur-rich surface. At these hot spots, the bulk atomic C/S ratio may be significantly higher.

The upper limits for carbon on Io overlap with the C/S ratios of ordinary chondrites. Geophysical models predict that Io has a Fe/Si ratio similar to that of the L or LL ordinary chondrites (Kuskov & Kronrod 2000, 2001), or to that of the CV3 or CM2 carbonaceous chondrites (Sohl et al. 2002). Cosmochemical models predict that Io accreted from CV3 or CM2 carbonaceous chondrite-like material (Consolmagno 1981; Lewis 1982; Prinn & Fegley 1981), which contains about 0.5–2.2 % carbon and 2.2–2.7 % sulfur by mass. For reference, the L and LL chondrites contain about 0.25–0.31 % carbon and 2.1–2.2 % sulfur by mass. It is plausible that the bulk C/S ratio of Io is similar to that of ordinary, CV3, or CM2 chondrites. However, some of the carbon and much of the sulfur in chondritic material would go into a metal-sulfide core, such as Io is believed to have (Anderson et al. 1996). Smaller amounts of C and S would be left in the silicate part of Io. Because of the uncertainties involved, we computed chemical equilibria for C/S ratios ranging from $10^{-6}$ to $10^{-1}$. This range spans the upper limits for carbon gases derived from observations, as can be seen in Fig. 1. We discuss the results of our chemical equilibrium calculations below.

## 4. RESULTS OF EQUILIBRIUM CALCULATIONS

### *4.1 Chemical reaction and eruption timescales*

Our thermodynamic calculations assume that Ionian volcanic gases reach chemical equilibrium inside the volcanic conduit where the high temperatures and pressures lead to characteristic chemical reaction times that are less than the characteristic eruption times ($t_{chem} < t_{erupt}$). However, the low temperatures and pressures in the volcanic plume lead to the opposite situation, namely ($t_{chem} > t_{erupt}$), and chemical equilibrium is not attained. In between these two regions, in the vicinity of the volcanic vent, $t_{chem}$ and $t_{erupt}$ become equal and quenching of the high temperature chemical equilibria occurs during the supersonic eruptions. Figure 1 of Zolotov

& Fegley (1999) illustrates our basic assumption. We earlier showed that chemical equilibrium is attained for sulfur and oxygen chemistry (Zolotov & Fegley 1998) and for alkali and halogen chemistry (Schaefer & Fegley 2004). Here we examine carbon chemistry.

The characteristic eruption time is 150–250 seconds (Zolotov & Fegley 1998). The characteristic chemical reaction times are defined as

$$t_{chem} = [i]/(d[i]/dt) \quad (1)$$

where [i] is the number density of species "i" given by its equilibrium mole fraction ($X_i$) times the total number density, and d[i]/dt is its destruction rate computed using the reaction rate constants given below. We considered a number of elementary reactions in the C–O–S system (Table 1), and found that reactions R1–R4 rapidly equilibrate at the nominal temperature (1760 K) and pressure (0.01 bar) of the Pele vent (see Table 2). However, reactions between CO and $CO_2$ such as R5–R7 are significantly slower than the characteristic eruption times of 150–250 seconds (Table 2). Reactions R5–R7 may not equilibrate unless they are catalyzed by either the magma and/or rock in the volcanic conduit, and/or other gas phase chemistry. This problem is analogous to that in the strato-mesosphere of Venus where $CO_2$ is reformed from CO and $O_2$ (Yung & DeMore 1999).

Catalysis of Ionian volcanic gas chemistry by the magma and/or rock is plausible but cannot be quantitatively modeled. We suggest that reactions R9–R12 in Table 2 may equilibrate CO and $CO_2$ in volcanic vents on Io. The net effect of reactions R9–R12 is

$$CO + O_2 \rightarrow CO_2 + O \quad (2)$$

This set of reactions is analogous to chlorine catalyzed chemistry proposed for Venus (catalytic cycle C19 in Appendix E) by Mills (1998). Alternatively, the well known reaction R8 could equilibrate CO and $CO_2$ in volcanic vents on Io, but the total hydrogen abundance required is very high (~0.1% of the sulfur abundance).

*4.2 Carbon equilibrium chemistry in Ionian volcanic gases*

We described the equilibrium chemistry of S, O, Na, Cl, K in our earlier papers (Fegley & Zolotov 2000a; Zolotov & Fegley 1999), and we focus on carbon equilibrium chemistry here. All carbon is in the gas, and no pure solid or molten carbon compounds are stable under the conditions we considered. Figure 2 shows the mole fractions[1] of the major carbon gases at our nominal conditions (T = 1760 ± 210 K, P = 0.01 bars) as a function of the C/S ratio. Our calculated mole fractions for CO and $CO_2$ are almost identical, with $CO_2$ being slightly larger. Carbonyl sulfide is ~2.5 orders of magnitude less abundant than CO, and $CS_2$ is ~3.5-4 orders of magnitude smaller than OCS. The abundances of all of the carbon gases increase with increasing C/S ratio, which is not surprising. Below are equations giving the mole fraction of each carbon gas as a function of bulk atomic C/S ratio in the gas at our nominal T and P:

$$\log\ X_{CO_2} = -0.27 + 0.99 \log\left(C/S\right) \tag{4}$$

$$\log\ X_{CO} = -0.33 + 1.00 \log\left(C/S\right) \tag{5}$$

$$\log\ X_{OCS} = -3.04 + 1.01 \log\left(C/S\right) \tag{6}$$

$$\log\ X_{CS_2} = -6.64 + 1.03 \log\left(C/S\right) \tag{7}$$

At our highest C/S ratio, where carbon is 10% of total sulfur, $CO_2$ has an abundance of ~8% relative to $SO_2$. At our lowest C/S ratio, where carbon is 0.01 % of total sulfur, $CO_2$ has an abundance of ~0.007% relative to $SO_2$.

Figure 2 also shows the upper limit for $CO_2$ measured by Voyager for the Loki plume (dotted line). This upper limit is not strictly applicable to the Pele hot spot, which we are using for our nominal temperature and pressure conditions. However, we cannot calculate chemical equilibria at Loki because the bulk composition of volcanic gas and the total pressure are unknown. The only data available for Loki are vent temperatures from Galileo and terrestrial observations, detections of $SO_2$ (Pearl et al. 1979) and SO (DePater et al. 2002), and the upper limits mentioned earlier. We can calculate chemical equilibria for Pele where sufficient data are available, but there are no upper limits (except for $CS_2$) on carbon gases. If we accept the upper limit for $CS_2$ as the actual abundance, the gas at Pele would have a bulk C/S atomic ratio > 10,000, which is geochemically implausible. Therefore, until more data become available, we apply the upper limits at Loki to Pele. Figure 2 shows that our calculated $CO_2$ abundance exceeds the upper limit for all C/S ratios greater than ~$10^{-3}$. We therefore use C/S = $10^{-3}$ for the nominal carbon abundance in our subsequent calculations.

[1] For reference, the mole fraction ($X_i$) of a gas is defined as equal to its partial pressure ($P_i$) divided by the total pressure ($P_T$), which is equivalent to the number of moles of a gas ($n_i$) divided by total moles ($n_T$) of all gases:

$$X_i = P_i/P_T = n_i/n_T \tag{3}$$

Figure 3 illustrates our results for carbon equilibrium chemistry as a function of temperature and total pressure. The calculations for Fig. 3a were performed at the nominal temperature of 1760 K as a function of variable pressure. Our nominal pressure is highlighted on the graph. Figure 3a shows that the abundances of $CO_2$ and OCS increase with increasing pressure, whereas the abundance of CO increases with decreasing pressure. At high total pressures, $CO_2$ is the most abundant gas followed by CO and OCS. The maximum abundance of OCS at our nominal temperature is only ~0.8% of $CO_2$ at a pressure of $10^2$ bars. At pressures greater than our nominal pressure, the $CO_2$ abundance exceeds the Voyager upper limit for the Loki plume. The $CS_2$ mole fraction is less than $10^{-8}$, and we do not expect $CS_2$ to be observable under any conditions on Io. We therefore believe that the upper limits for $CS_2$ from McGrath et al. (2000) are much higher than the actual abundances.

Figure 3b shows carbon equilibrium chemistry at our constant nominal pressure for a range of temperatures. Our nominal temperature is highlighted on the graph. At temperatures above our nominal temperature, CO is the major gas, followed by $CO_2$, OCS and $CS_2$. The CO abundance decreases with decreasing temperature, whereas that of $CO_2$ increases with decreasing temperature. At temperatures below our nominal value of 1760 K, the abundance of $CO_2$ exceeds the upper limit determined by Voyager for the Loki plume. Loki is apparently a lava lake that periodically overturns (Rathbun et al. 2002). It typically has temperatures of 400 K (Geissler 2003), much less than those of Pele, but peak temperatures at Loki exceed 900 K (Rathbun et al 2002). We can get $CO_2$ abundances less than the Voyager upper limits, if we decrease (1) the pressure, (2) the C/S ratio, or (3) both. Carbonyl sulfide has a fairly constant abundance from 2000 K down to about 600 K, where the abundance drops sharply. The $CS_2$ mole fraction is less than $10^{-8}$ and is not shown on the graph.

Figure 4 illustrates the effect of the O/S molar ratio on carbon equilibrium chemistry. This ratio is the bulk O/S ratio of the volcanic gas. Pure $SO_2$ has O/S = 2, pure $SO_3$ has O/S = 3, and pure sulfur vapor has O/S equal to zero. Oxygen to sulfur ratios from 0–2 can be thought of as mixtures of $SO_2 + S_2$, whereas O/S ratios from 2–3 can be thought of as mixtures of $SO_2 + O_2$. To date, there is no evidence for O/S ratios >2 in volcanic gases on Io. The oxygen fugacity ($fO_2$, essentially the $O_2$ partial pressure) varies with temperature, total pressure, and the O/S molar ratio of the volcanic gas (Zolotov & Fegley 1999). The range of $fO_2$ values as a function of the O/S ratio at constant T (1760 K), P (0.01 bar), and C/S ratio ($10^{-3}$) is shown along the top of Fig.

4. Carbon dioxide becomes more abundant, while CO, OCS, and $CS_2$ are oxidized by $O_2$ and become less abundant as the $fO_2$ increases, e.g.,

$$2\ CO + O_2 = 2\ CO_2 \quad (8)$$

$$2\ OCS + 3\ O_2 = 2\ CO_2 + 2SO_2 \quad (9)$$

$$CS_2 + O_2 = 2\ CO_2 + 2SO_2 \quad (10)$$

Figure 4 shows CO and $CO_2$ are about equally abundant at the $fO_2$ for Pele volcanic gases. As we discussed previously, Loki is believed to be more oxidizing than Pele (Zolotov & Fegley 1999). Thus if all other variables were the same, the $CO_2$/CO ratio at Loki would be larger than that at Pele. The lower temperature at Loki would also tend to increase the $CO_2$/CO ratio as shown in Fig. 3b. However, neither the pressure nor bulk O/S ratio are known so their effects on the $CO_2$/CO ratio at Loki cannot be quantified at this time.

## 5. DISCUSSION

### *5.1 Evidence for the presence of carbon on Io*

Garrard et al. (1996) describe observations of Jupiter's magnetosphere by the Galileo heavy ion counter (HIC). They detected highly ionized solar C which had variable abundances depending upon energy. The abundances outside of Io's orbit (C/O = 0.026±0.007 at 16-17MeV per nucleon) were significantly larger than the abundances inside of Io's orbit (C/O ~ 0.007 ± 0.003 at 16 to 17 MeV). Their measured S/O ratios were ~1.2 ±0.2 over all energy ranges, both inside and outside of Io's orbit. This gives C/S ratios of ~0.022 outside of Io's orbit and ~0.0058 inside of Io's orbit. The upper limits for the $C^{2+}/S^{2+}$ and $C^{+}/S^{2+}$ ratios in the IPT are smaller than these values and are $< 3.7\times10^{-4}$ and $< 2.5\times10^{-3}$, respectively (Feldman et al. 2004). The two sets of observations likely give different ratios due to a sampling bias: the HIC was observing highly ionized solar wind particles, whereas the observations of Feldman and colleagues are sampling Iogenic, low energy particles. However, the lower C/S ratio inside of Io's orbit observed by the HIC indicates that Io absorbs carbon atoms at these energies. Therefore, it appears that Io has some limited amount of carbon absorbed from the solar wind. If this carbon sticks to the surface, then it may join the cycle of burial and recycling which has been proposed to take place (Keszthelyi et al. 2004). It would be very interesting to determine how much C Io is absorbing; however, such a calculation is outside of the scope of this paper.

*5.2 Photochemical lifetimes of carbon gases*

We now consider how carbon-bearing gases are destroyed by photochemistry in the volcanic plume. Table 3 lists the photochemical lifetimes of CO, $CO_2$, OCS, and $CS_2$ at Io. We calculated the photochemical lifetimes (at zero optical depth) from the $J_1$ values at 1 AU given in Appendix 1 of Levine (1985), scaled to the distance of Io from the Sun, where:

$$t_{chem} = \frac{1}{J_1} \qquad (11)$$

Table 3 shows that the lifetimes of all carbon gases are long compared to characteristic time scales of ~17 minutes for gas flow over 100 km on Io's surface and ~20 minutes for a ballistic lifetime within a large plume (see section 6 of Moses et al. 2002). The photochemical lifetimes of CO and $CO_2$ are also long compared to an Ionian day (~42 hours).

*5.3 Condensation of carbon gases on Io's surface*

Another possible loss mechanism for carbon gases is condensation onto Io's surface. To determine whether any of the carbon gases may condense on Io, we looked at the vapor pressures of these ices. It is unlikely that $CS_2$ is abundant enough to condense, and CO is too volatile to condense at Io's surface temperatures. The vapor pressure data for $CO_2$ and OCS from Honig & Hook (1960) at the surface temperatures of Io (90-120 K, Rathbun et al. 2004) give vapor pressure ranges of 7.58 – 45,586 nbars for $CO_2$, and 63 – 68,433 nbars for OCS. The atmosphere of Io, which is primarily $SO_2$, has a total pressure of 1 – 10 nanobars. The atmospheric partial pressure of a gas $P_i$ (atm) has to equal its vapor pressure over the solid ice $P_i$ (ice) for the frozen gas to be stable:

$$P_i\,(atm) = P_i\,(ice) \qquad (12)$$

Otherwise the ice is unstable and all the gas resides in the atmosphere. Therefore, OCS ice is unstable on the surface of Io because the atmospheric partial pressure of OCS is less than the vapor pressure over OCS ice. Carbon dioxide ice could possibly be condensed in colder areas on Io's surface (T < 86 K, Rathbun et al. 2004). Sandford et al. (1991) suggested that a band at 4705.2 $cm^{-1}$ was caused by clusters of $CO_2$ molecules, which could be present on the surface in $SO_2$ or in the atmosphere as aerosol particles. However, they found that the entire surface of Io would have to be dominated by tetramer clusters of $CO_2$ in order to match both the width and location of the band. Schmitt et al. (1994) instead assigned the same band to an $SO_2$ overtone. We conclude that condensation is unlikely to be a significant loss mechanism for carbon-bearing gases (except possibly $CO_2$) on Io.

### *5.4 Chemical destruction of carbon gases*

Photochemical modeling of Io's atmosphere and volcanic plumes (Moses et al. 2002 and references therein) indicates that the most abundant reactive species are O, S, SO, and $O_2$. We looked at a number of reactions of these species with CO, $CO_2$, OCS, and $CS_2$ to determine which, if any, chemical reactions could destroy these carbon gases (see Table 1). We calculated chemical lifetimes for CO, $CO_2$, OCS, and $CS_2$ assuming surface conditions similar to those used in Moses et al. (2002) of 110 K, $P_T$ = 10 nbars, and a total number density of $10^{11.82}$ $cm^{-3}$. All of these reactions give chemical lifetimes (Table 4) much longer than the photochemical lifetimes (Table 3). The fastest of these reactions is R4, which gives a chemical lifetime for $CS_2$ of ~146 days. This is very large compared with the photochemical lifetime of ~58 minutes for $CS_2$. We conclude that photolysis is probably the primary loss mechanism for $CO_2$, CO, OCS, and $CS_2$ in volcanic plumes and Io's atmosphere.

## 6. CONCLUSIONS

Chemical equilibrium calculations predict that CO and $CO_2$ are the two most important carbon compounds in volcanic gases on Io. Carbonyl sulfide and $CS_2$ are much less abundant and will be difficult, if not impossible to observe. The upper limits reviewed in Figure 1 and our calculations support an average upper limit for the bulk atomic C/S ratio < $10^{-3}$ for volcanic gases on Io. Using this bulk C/S ratio we predict $CO/SO_2 < 6\times10^{-4}$ and $CO_2/SO_2 < 8\times10^{-4}$ at Pele. Our considerations of photolysis, condensation, and disequilibrium reactions indicate that photolysis is probably the major loss process for CO, $CO_2$, OCS, and $CS_2$. The atmospheric partial pressures of these gases are too low for their pure ices to be stable on Io's surface; however, carbon dioxide may condense in cold traps on the surface. Their long photochemical lifetimes indicate that CO and $CO_2$ may be present in Io's atmosphere at all times. However, CO will most likely be observable during high temperature (> 1500 K) volcanic eruptions. Carbon dioxide will be most abundant in low temperature/high pressure eruptions. It may be possible to observe OCS in a volcanic plume during an active eruption with a sufficiently high eruption pressure. A detection of $CS_2$ gas would indicate a lava with a very high C/S ratio. We particularly recommend searching for $CO_2$, especially near active volcanic vents such as Pele, where its abundance should be greater. We also recommend searching for CO during high temperature eruptions, when it should be more abundant.

## ACKNOWLEDGMENTS

We thank K. Lodders and M. McGrath for helpful comments and J. Spencer for a constructive review and helpful comments. This work was supported by the NASA Planetary Atmospheres Program.

Table 1
Kinetics of Chemical Reactions Involving Carbon Gases

| No. | Reaction | Rate Constant[a] | Reference |
|---|---|---|---|
| R1 | $OCS + O \rightarrow CO + SO$ | $2.09\times10^{-11}\exp(-2200/T)$ | 1 |
| R2 | $OCS + S \rightarrow S_2 + CO$ | $1.52\times10^{-12}\exp(-1827/T)$ | 2 |
| R3 | $CO + SO_2 \rightarrow$ products | $4.47\times10^{-12}\exp(-24301/T)$ | 3 |
| R4 | $CS_2 + O \rightarrow CS + SO$ | $3.20\times10^{-11}\exp(-650/T)$ | 1 |
| R5 | $CO + O + M \rightarrow CO_2 + M$ | $1.7\times10^{-33}\exp(-1510/T)$ | 4 |
| R6 | $CO + O_2 \rightarrow CO_2 + O$ | $4.2\times10^{-12}\exp(-24{,}000/T)$ | 4 |
| R7 | $CO_2 + O \rightarrow CO + O_2$ | $2.8\times10^{-11}\exp(-26{,}500/T)$ | 4 |
| R8 | $CO + OH \rightarrow CO_2 + H$ | $3.52\times10^{-12}\exp(-2630/T)$ | 5 |
| R9 | $Cl + CO + M \rightarrow ClCO + M$ | $3.28\times10^{-24}\,T^{-3.8}$ | 1 |
| R10 | $ClCO + O_2 + M \rightarrow ClCO_3 + M$ | $\frac{5.7\times10^{-15}\exp(500/T)}{(1\times10^{17}+0.05M)}$ | 6 |
| R11 | $ClCO_3 + SO \rightarrow Cl + CO_2 + SO_2$ | $1.0\times10^{-11}$ | 6 |
| R12 | $SO_2 + M \rightarrow SO + O$ | $6.61\times10^{-9}\exp(-50{,}517/T)$ | 7 |

[a]Two-body rate constants are in units of $cm^3\ s^{-1}$. Three-body reactions are in units of $cm^6\ s^{-1}$. M is any third body.
References.—(1) DeMore et al. 1997 (2) Klemm & Davis 1974, (3) Bauer et al. 1971, (4) Tsang & Hampson 1986, (5) Wooldridge et al. 1994, (6) Mills 1998, (7) Plach & Troe 1984.

Table 2
Chemical Lifetimes of Carbon Gases at the Pele Vent (T = 1760 K, P = $10^{-2}$ bar)

| Reaction | Species Destroyed | $t_{chem}$ (s) |
|---|---|---|
| R1 | OCS | $\sim 3.00\times10^{0}$ |
| R2 | OCS | $\sim 3.00\times10^{-2}$ |
| R3 | CO | $\sim 7.00\times10^{0}$ |
| R4 | $CS_2$ | $\sim 8.00\times10^{-1}$ |
| R5 | CO | $\sim 7.94\times10^{5}$ |
| R6 | CO | $\sim 1.58\times10^{6}$ |
| R7 | $CO_2$ | $\sim 2.00\times10^{6}$ |

Table 3
Photochemical Lifetime of Carbon Gases in the Atmosphere of Io

| Species | $J_1$ ($s^{-1}$) | $t_{chem}$ (s) |
|---|---|---|
| CO[a] | $2.4\times10^{-8}$ | $4.17\times10^{7}$ |
| $CO_2$ | $7.40\times10^{-8}$ | $1.35\times10^{7}$ |
| OCS | $2.40\times10^{-5}$ | $4.17\times10^{4}$ |
| $CS_2$ | $1.29\times10^{-4}$ | $3.47\times10^{3}$ |

[a]The $J_1$ value for CO is for the combined dissociation reactions: (a) C + O; (b) C($^1$D) + O($^1$D); (c) $CO^+ + e^-$; (d) O + $C^+$ + e-; (e) C + $O^+$ + e-.
References.— Levine (1985)

Table 4
Chemical Lifetimes of Carbon Gases Inside the Pele Plume (T = 110 K, P = $10^{-8}$ bar)

| Reaction | Species Destroyed | $t_{chem}$ (s) |
|---|---|---|
| R1 | OCS | $2.51\times10^{13}$ |
| R2 | OCS | $9.33\times10^{9}$ |
| R3 | CO | $3.98\times10^{95}$ |
| R4 | $CS_2$ | $1.26\times10^{7}$ |
| R5 | CO | $1.00\times10^{21}$ |
| R6 | CO | $7.94\times10^{99}$ |
| R7 | $CO_2$ | $1.00\times10^{109}$ |
| R9 | CO | $1.58\times10^{12}$ |

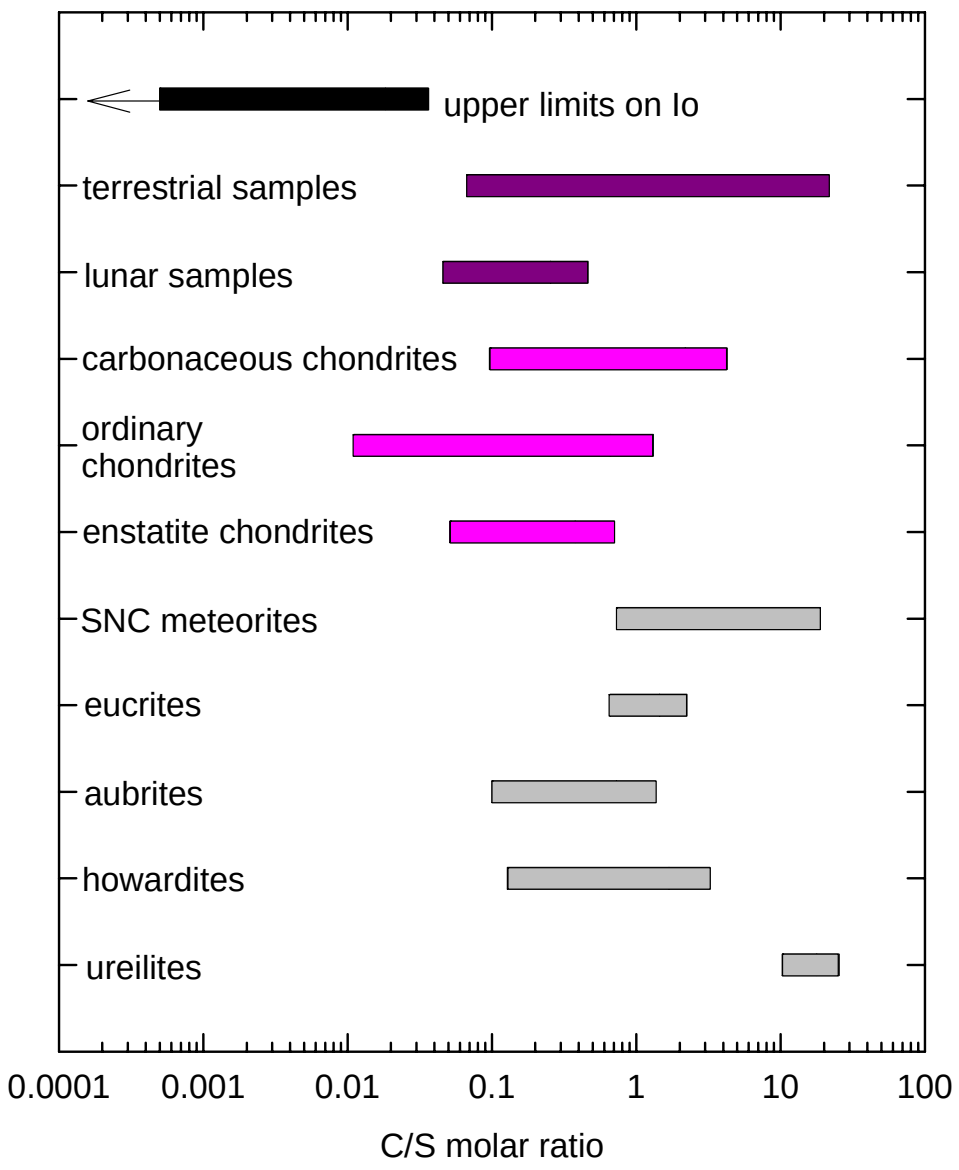

Figure 1: C/S atomic ratios in lunar, meteoritic, and terrestrial samples are compared to the upper limits for C/S in Io's atmosphere and the Io plasma torus. The upper bound on the range for Io is the upper limit for $CH_4$ (g) in Loki's volcanic plume determined by Voyager. The lower bound is the upper limit for $CO_2$ from the same observation. Data sources for the lunar, meteoritic, and terrestrial samples and the types of samples are listed in the text.

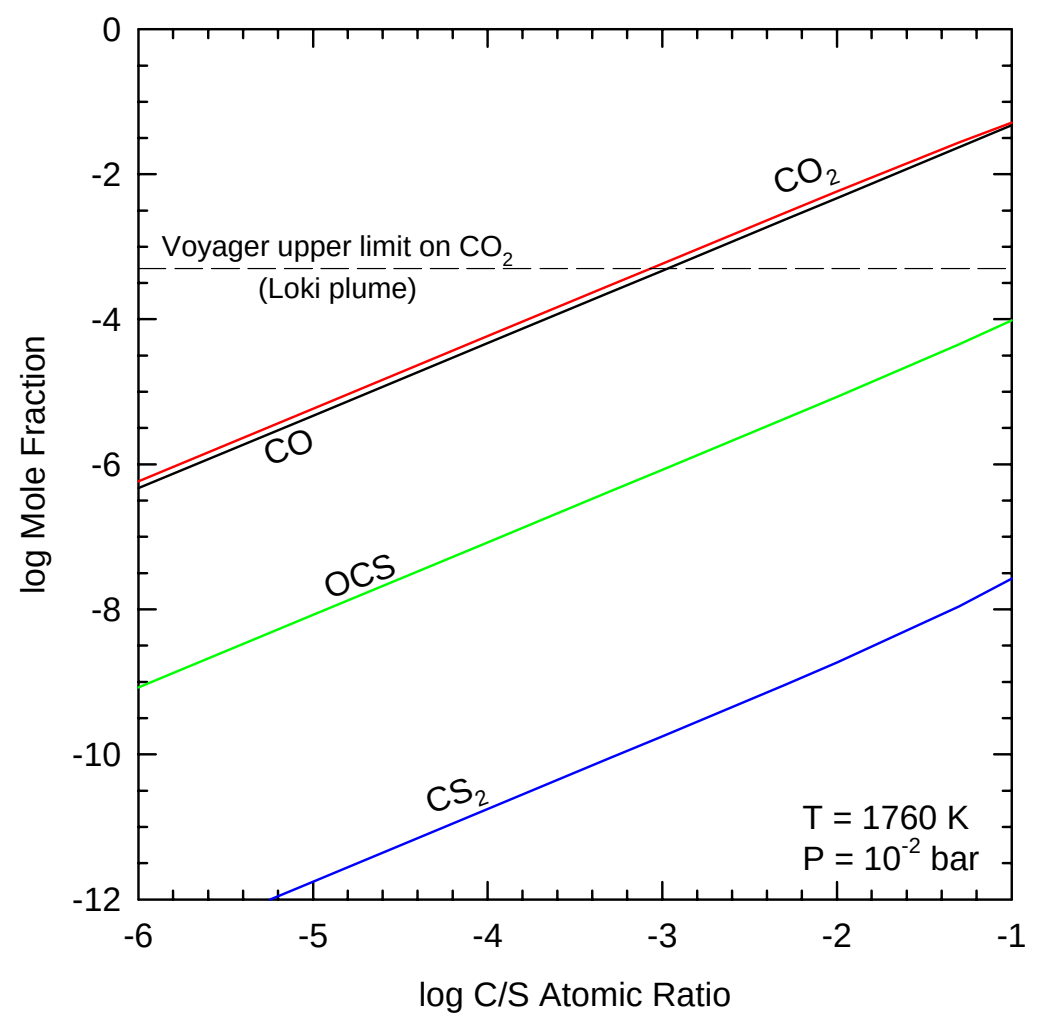

Figure 2: Carbon gas chemistry as a function of the C/S atomic ratio at the nominal conditions for the Pele volcanic vent (T = 1760 K, P = 0.01 bars). The shaded region shows an uncertainty of ±210 K from Lopes et al. (2001).

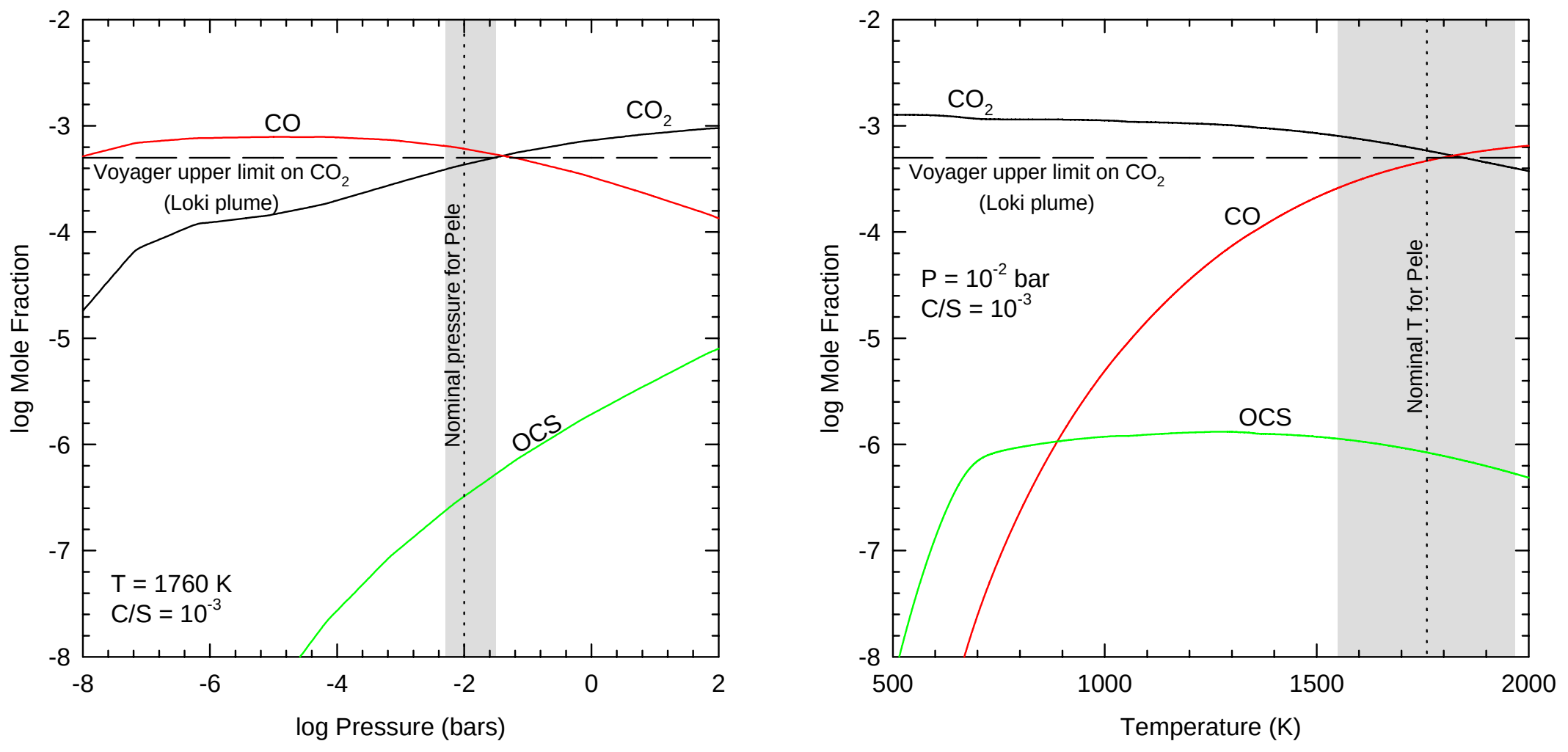

Figure 3: Equilibrium chemistry of carbon in a volcanic gas containing C, S, O, Na, K, and Cl for a C/S ratio = 0.001 (a) at a constant temperature of 1760 K as a function of pressure and (b) at a constant pressure of 0.01 bars as a function of temperature. The shaded region in part (a) shows the nominal Pele pressure plus uncertainties, and the shaded region in part (b) shows the nominal Pele temperature, plus uncertainties.

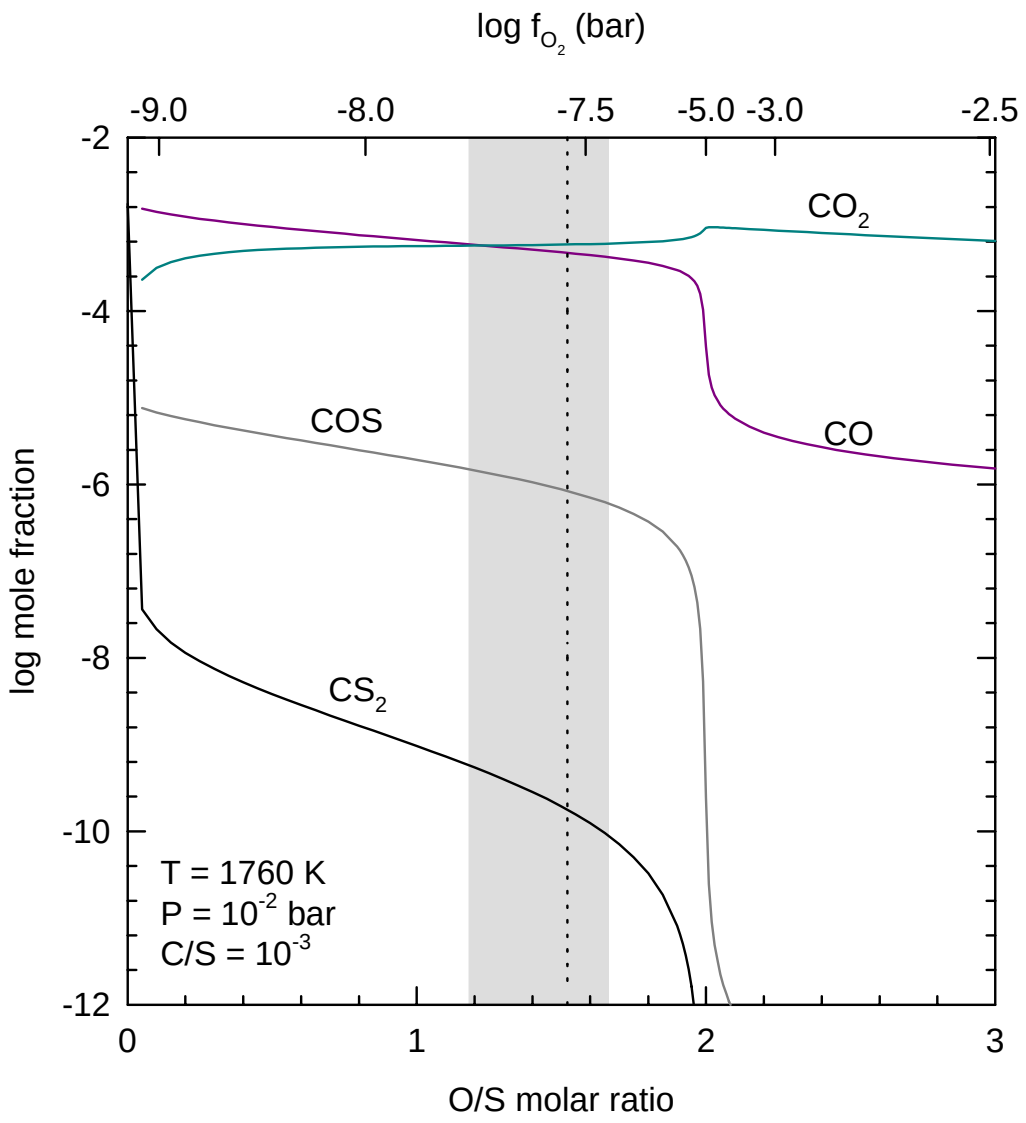

Figure 4. Equilibrium chemistry of carbon in the same volcanic gas as a function of the O/S ratio at otherwise constant conditions (T = 1760 K, P = 0.01 bar, C/S = 0.001).